\documentclass[12pt]{article}
\usepackage{graphicx}
\usepackage{amssymb}
\usepackage{amsmath} %arrows
\usepackage{multirow} 
\usepackage{rotating}
\usepackage{array}
\usepackage[margin=2cm]{geometry}
\usepackage{cite}
\usepackage[center,footnotesize,hang]{subfigure}
\usepackage{amsfonts}
\usepackage[footnotesize]{caption}
\usepackage{url}
\usepackage{relsize}
\usepackage{makecell}
\usepackage{fullpage}

\newcommand{\pmatr}[1]{\begin{pmatrix} #1 \end{pmatrix}}

\def\be{\begin{equation}}
\def\ee{\end{equation}}
\def\beq{\begin{equation}}
\def\eeq{\end{equation}}
\def\bea{\begin{eqnarray}}
\def\eea{\end{eqnarray}}

\def\be{\begin{equation}}
\def\ee{\end{equation}}
\def\beq{\begin{equation}}
\def\eeq{\end{equation}}
\def\bea{\begin{eqnarray}}
\def\eea{\end{eqnarray}}

\newcommand{\newc}{\newcommand}
\newc{\ol}{\overline}
\newc{\wt}{\widetilde}
\newc{\bs}{\boldsymbol}
\newc{\ma}{\mathcal}
\newc{\vl}{\langle}
\newc{\vr}{\rangle}

\newc{\sg}{S}
\newc{\ug}{U}
\newc{\tg}{T}

\begin{document}

\title{\hfill ~\\[-30mm]
          \hfill\mbox{\small  QFET-2016-12}\\[-3.5mm]
          \hfill\mbox{\small  SI-HEP-2016-20}\\[13mm]
       \textbf{
       Littlest Seesaw model from  $\boldsymbol{S_4 \times U(1)}$ 
}}

\author{\\[-3mm]
Stephen F. King$\,^a\,$\footnote{E-mail: {\tt king@soton.ac.uk}}
,~~ Christoph Luhn$\,^b\,$\footnote{E-mail: {\tt christoph.luhn@uni-siegen.de}}
\\[8mm]
$^a$\,\it{\small School of Physics and Astronomy, University of Southampton,}\\
\it{\small SO17 1BJ Southampton, United Kingdom}\\[2mm]
$^b$\,\it{\small Theoretische Physik 1, Naturwissenschaftlich-Technische
    Fakult\"at, Universit\"at Siegen,}\\
\it{\small Walter-Flex-Stra{\ss}e 3, 57068 Siegen, Germany}
}

\date{}

\maketitle

\begin{abstract}
\noindent  
We show how a minimal (littlest) seesaw model involving two right-handed neutrinos and a very constrained Dirac mass matrix, with one texture zero and two independent Dirac masses, may arise from $S_4\times U(1)$ symmetry in a semi-direct supersymmetric model. The resulting CSD3 form of neutrino mass matrix only depends on two real mass parameters plus one undetermined phase. We show how the phase may be fixed to be one of the cube roots of unity by extending the $S_4\times U(1)$ symmetry to include a product of $Z_3$ factors together with a CP symmetry, which is spontaneously broken leaving a single residual $Z_3$ in the charged lepton sector and a residual $Z_2$ in the neutrino sector, with suppressed higher order corrections. With the phase chosen from the cube roots of unity to be $-2\pi/3$, the model predicts a normal neutrino mass hierarchy with $m_1=0$, reactor angle $\theta_{13}=8.7^\circ$, solar angle $\theta_{12}=34^\circ$, atmospheric angle $\theta_{23}=44^\circ$, and CP violating oscillation phase $\delta_{\rm CP}=-93^\circ$, depending on the fit of the model to the neutrino masses.

\end{abstract}
\thispagestyle{empty}
\vfill

\newpage

\setcounter{page}{1}

\section{Introduction}

Despite great experimental progress in neutrino physics in the last twenty
years~\cite{nobel}, the origin of neutrino mass and lepton mixing remains unclear. 
Although there has been intense theoretical activity in this period, there is
still no leading candidate for a theory of neutrino mass and lepton mixing
(for reviews see e.g.~\cite{King:2013eh,King:2015aea}).

From a theoretical point of view the most appealing possibility seems to be
the seesaw mechanism in its original formulation involving heavy right-handed
Majorana neutrinos~\cite{seesaw}. However the seesaw mechanism is very
difficult to test experimentally, at least if the right-handed neutrino masses
are beyond reach of the LHC, and also introduces many additional
parameters. One approach to this problem is to follow the idea of minimality,
leading to seesaw theories with smaller numbers of parameters and hence
testable predictions~\cite{King:2015sfk}. If the predictions are realised
experimentally then this may provide indirect experimental support for the
seesaw mechanism, and in addition provide insights into the flavour
problem. This is the approach we shall follow in this paper.

The most minimal version of the seesaw mechanism involves two right-handed
neutrinos~\cite{King:1999mb}. In order to reduce the number of free parameters
still further to the smallest number possible, and hence increase predictivity, various
approaches to the two right-handed neutrino seesaw model have been suggested,
such as postulating one~\cite{King:2002nf} or two~\cite{Frampton:2002qc}
texture zeroes, however such two texture zero models are now
phenomenologically excluded~\cite{Harigaya:2012bw} for the case of a normal
neutrino mass hierarchy considered here. The minimal successful scheme with
normal hierarchy seems to be a two right-handed model with a Dirac mass matrix
(in the diagonal charged lepton mass basis) involving one texture zero and a
particular pattern of couplings, together with a diagonal right-handed
neutrino mass matrix~\cite{King:2013iva}, 
\begin{equation}
	m^D = \pmatr{0 & b \\ a & 3b \\ a & b } , \ \ \ \ 
	M_{R}=
\left( \begin{array}{cc}
M_{\rm atm} & 0 \\
0 & M_{\rm sol}
\end{array}
\right),
	\label{mD}
\end{equation}
where $a,b$ are two complex parameters. The seesaw
mechanism~\cite{seesaw} leads to a light effective Majorana neutrino mass
matrix: 
\begin{equation}
	m^\nu = m_a 
	\left(
\begin{array}{ccc}
	0&0&0\\0&1&1\\0&1&1 
	\end{array}
\right)
	+ m_b e^{i\eta } 
	\left(
\begin{array}{ccc}
	1&3&1\\3&9&3\\1&3&1
	\end{array}
\right) .
	\label{eq:mnu2p100}
\end{equation}
$m_a= |a|^2/M_{\rm atm}$ and $m_b= |b|^2/M_{\rm sol}$ may be taken to be real
and positive without loss of generality, the physical predictions only
depending on a relative phase whose phenomenologically preferred value is 
$\eta = 2\pi/3$~\cite{King:2013iva}. 
Following the proposed lepton model in~\cite{King:2013iva}, this structure
has been incorporated into unified models of quarks and leptons
in~\cite{Bjorkeroth:2015ora}. It has also been shown to lead to successful leptogenesis
in which not only the sign of baryon asymmetry is determined by the ordering of the heavy right-handed
neutrinos, but also $\eta$ is identified as the leptogenesis phase, directly linking CP violation in the laboratory with that in the early universe 
\cite{Bjorkeroth:2015tsa}.

The implementation of the seesaw mechanism above
is an example of sequential dominance (SD)~\cite{King:1998jw} 
in which the first term in Eq.~\eqref{eq:mnu2p100}, arising from the first 
(atmospheric) right-handed neutrino, provides the dominant contribution to the
atmospheric neutrino mass, leading to approximately maximal atmospheric
mixing, while the second subdominant term from the second (solar) right-handed
neutrino gives the solar neutrino mass and controls the solar and reactor
mixing and CP violation. If the constrained form of 
Dirac mass matrix in Eq.~\eqref{mD} is relaxed, but the texture zero is
maintained, then SD generally leads to a reactor angle which is bounded by
$\theta_{13}\lesssim m_2/m_3$~\cite{King:2002nf}, a prediction 
that was made a decade before the reactor angle was measured in
2012~\cite{nobel}. However sharp predictions for the reactor (and solar)
angles can only result from applying constraints to the Dirac mass matrix of
various types, an approach known as constrained sequential dominance
(CSD)~\cite{King:2005bj}. For example, keeping the first column of the Dirac
mass matrix fixed $(0,a,a)^T$, a class of CSD$n$ models has emerged~\cite{King:2005bj,Antusch:2011ic,King:2013iva,King:2013xba,Bjorkeroth:2014vha} 
corresponding to the second column taking the form $(b,nb,(n-2)b)^T$, with a
reactor angle approximately given by~\cite{King:2015dvf} 
\beq
\theta_{13} \sim (n-1) \frac{\sqrt{2}}{3}  \frac{m_2}{m_3},
\label{s130}
\eeq
where CSD1~\cite{King:2005bj} implies tri-bimaximal (TB) mixing with a zero
reactor angle, CSD2~\cite{Antusch:2011ic} has a reactor angle 
$\theta_{13} \sim \frac{\sqrt{2}}{3}  \frac{m_2}{m_3}$, which is too small,
CSD3~\cite{King:2013iva} in Eq.~\eqref{mD} predicts $\theta_{13} \sim
\frac{2\sqrt{2}}{3}  \frac{m_2}{m_3}$ which is in good agreement with the
experimental value $\theta_{13}\sim 0.15$~\cite{nobel},
and CSD4~\cite{King:2013xba} predicts $\theta_{13} \sim
\sqrt{2}  \frac{m_2}{m_3}$, while higher values of $n> 4$ involve increasingly
large values of the reactor angle which are disfavoured~\cite{Bjorkeroth:2014vha}.

The seesaw scheme in Eq.~\eqref{mD} is referred to as either CSD3 or the Littlest
Seesaw (LS)~\cite{King:2015dvf} since the seesaw mechanism only involves two
complex Dirac masses $a,b$ together with two real positive right-handed
neutrino masses $M_{\rm atm}$ and $M_{\rm sol}$ (as compared to 18 parameters
in the most general three right-handed neutrino seesaw mechanism). 
The resulting neutrino mass matrix in Eq.~\eqref{eq:mnu2p100} involves only three
parameters, namely the real positive mass parameters $m_a, m_b$ together with
the real phase $\eta$. It was realised~\cite{King:2013iva} that if the phase
is also fixed to be $\eta = 2\pi/3 $ then this leads to a highly predictive and
successful scheme, with only two remaining real positive input parameters
$m_a,m_b$ which may be determined by the physical neutrino masses $m_2,m_3$,
with $m_1=0$ being an automatic prediction of two right-handed neutrinos. The entire
PMNS mixing matrix is then uniquely predicted by the model.

Although the Littlest Seesaw is unquestionably minimal and predictive, the
{\it Achilles Heel} of this model has always been its theoretical
justification from symmetry. For example, assuming some family symmetry,
spontaneously broken by some new Higgs fields (the so-called flavons) in the
triplet representation, the structure of the Dirac mass matrix in
Eq.~\eqref{mD} may in principle arise from the vacuum alignment of these
flavons. However, the desired flavon vacuum alignment $(1,3,1)^T$, responsible
for the second column of the Dirac mass matrix, does not seem to follow
directly from any symmetry, but only indirectly via a sequence of flavon
alignments which are mutually orthogonal~\cite{King:2013iva,Bjorkeroth:2014vha}. 
However it was recently realised that $S_4$ might be the best candidate
symmetry for producing this alignment~\cite{King:2015dvf} since in the real
basis it is the minimal symmetry that preserves a $U$ type symmetry capable of
equating two of the elements of the alignment, namely the first and third
components of $(1,3,1)^T$. However to date it has not proved possible to
construct a model in which both the neutrino mass matrix  and charged lepton
mass matrix structures are enforced by subgroups of the original family
symmetry.\footnote{In general, neutrino mass models based on
discrete family symmetry may be classified into three
types~\cite{King:2009ap}: direct, semi-direct and indirect, depending on the
residual symmetry preserved in the neutrino and charged lepton sectors. If the
full Klein symmetry of the Majorana neutrino mass matrix and the symmetry of
the charged lepton mass matrix are identified as subgroups of the original
family symmetry, the models are known as direct, while semi-direct (or
indirect) models correspond to cases where only a part (or none) of the
residual symmetries may be identified as subgroups of the family symmetry.}

In this paper, then, we shall propose a Littlest Seesaw model in which a minimal 
neutrino mass matrix, simply related to that in Eq.~\eqref{eq:mnu2p100}, follows 
from a semi-direct supersymmetric model plus some minimal dynamical constraints.
This represents real progress since previously the Littlest Seesaw has only
been realised in indirect models not enforced by any (discrete)
symmetry considerations.  In our semi-direct approach here we shall
use $S_4 \times U(1)$ to enforce a version of the Littlest Seesaw which 
is simply related to that in Eqs.~(\ref{mD},\ref{eq:mnu2p100}),
by the permutation $L_2 \leftrightarrow L_3$. We shall also show that this
new version of CSD3 may also be generalised to CSD$n$.
The starting point for our approach here is the observation 
that Eq.~\eqref{eq:mnu2p100} leads to trimaximal~TM$_1$
mixing~\cite{Xing:2006ms,Albright:2008rp}, 
in which the first column of the tri-bimaximal
mixing matrix~\cite{Harrison:2002er} is preserved.
The inspiration for our approach comes from the semi-direct model of trimaximal~TM$_1$ mixing that was developed in~\cite{Luhn:2013vna} in which, 
denoting the three generators of $S_4$ as $S,U,T$, the model preserves a
residual $Z_3$ in the charged lepton sector arising from the $T$ generator,
and a $Z_2$ in the neutrino sector corresponding to the product $SU$.
Following~\cite{Luhn:2013vna}, we shall enforce the Littlest Seesaw by similar 
symmetry arguments, the notable difference being that in our case, instead of having 
three right-handed neutrinos in a triplet of $S_4$, the model here involves 
two right-handed neutrinos which are singlets of $S_4$. 

We shall also impose a CP symmetry in the original theory which is
spontaneously broken, where unlike~\cite{Ding:2013hpa},
there is no residual CP symmetry in either the charged lepton
or neutrino sectors. Nevertheless we shall obtain sharp predictions for CP
violation by fixing the phase~$\eta$ in the neutrino mass matrix
Eq.~\eqref{eq:mnu2p100} to be one of the cube roots of unity due to a $Z_3$
family symmetry, using the mechanism proposed in~\cite{Antusch:2011sx}.
In order to achieve this, we suppose that the original $U(1)$
which accompanies $S_4$ is extended to a product of $U(1)$ factors, where some
of these are supposed to be explicity broken to $Z_3$ subgroups, which are
subsequently spontaneously broken along with the $S_4$. This is perhaps the
least appealing feature of our scheme, but it is necessary in order to obtain
a sharp input value for the phase~$\eta$, and hence CP violation, as well as the lepton
mixing angles which also depend on~$\eta$. We shall propose a concrete models
along these lines based on $S_4$ together with one $U(1)$ factor accompanied
by five $Z_3$ symmetries, and show that the desired leading order operator
structure in both the Yukawa and vacuum alignment sectors have quite
suppressed higher order corrections, leading to reliable predictions for
observable neutrino masses as well as lepton mixing and CP parameters.

The layout of the remainder of the paper is as follows.
In Section~\ref{littlest} we show how the Littlest Seesaw can arise from $S_4$
symmetry, avoiding any technical details, making the paper accessible to any
casual reader. In Section~\ref{sec:align} we show how the necessary vacuum
alignments of CSD3 can arise from an $F$-term mechanism which does not rely on
long chains of orthogonality conditions and is simpler than previous attempts. In
Section~\ref{concrete} we describe a model of leptons based on $S_4\times
U(1)$ that leads to the Littlest Seesaw, then extend it to $S_4\times
U(1)\times (Z_3)^5$ in order to fix the phase to be a cube root of unity.
Finally in Section~\ref{CLFV} we briefly comment on charged lepton flavour
violation in this model.
Section~\ref{conclusions} concludes the paper. 
In addition, Appendix~\ref{app:CGs} gives the necessary group theory of $S_4$,
along with the symmetry preserved and broken by various vacuum alignments, and
the $S_4$ Clebsch-Gordan coefficients. Appendix~\ref{analytic} generalises the
version of CSD3 discussed in this paper to a new type of CSD$n$ and presents
analytic formulas for neutrino masses and lepton mixing parameters for this case.

%%%%%%%%%%%%%%%%%%%%%%%%%%%%%%%%%%%%%%%%%%%%%

%%%%%%%%%%%%%%%%%%%%%%%%%%%%%%%%%%%%%%%%%%%%%

%%%%%%%%%%%%%%%%%%%%%%%%%%%%%%%%%%%%%%%%%%%%%

\section{Littlest Seesaw model from $\boldsymbol{S_4}$: an overview}
\label{littlest}

Before getting into too many technicalities of symmetry and model building, it
is useful to give a sketch of the type of model we will present in this paper.
This enables serious readers to have in mind where we are heading before getting 
immersed in the details, or casual readers to simply read this section of the paper,
then jump to the Conclusions.
The version of the Littlest Seesaw model in this paper involves lepton doublets
which transform under $S_4$ as $L\sim {\bf 3}^{{\prime}}$, two right-handed
neutrinos $N^c_{\rm sol}\sim {\bf 1}$, $N^c_{\rm atm}\sim {\bf 1}$ and the up-
and down-type Higgs fields $H_{u,d}\sim {\bf 1}$ with couplings in the superpotential:
\be
\frac{\phi'_{\rm atm}}{\Lambda}LH_uN^c_{\rm atm}  
+ \frac{\phi'_{\rm sol}}{\Lambda}LH_uN^c_{\rm sol}  \ ,
\ee
where the non-renormalisable terms are suppressed by a dimensionful cut-off
$\Lambda$ and  the flavons $\phi'_{\rm atm}\sim {\bf 3}^{{\prime}}$ and
$\phi'_{\rm sol}\sim {\bf 3}^{{\prime}}$  are required to have the vacuum
alignments\footnote{The minus signs in the third components are related to the
$S_4$ triplet basis as defined in Appendix~\ref{app:CGs}.}
\be
\langle \phi'_{\rm atm} \rangle =\varphi'_{\rm atm}
\begin{pmatrix}
 0 \\
 1\\
 -1
\end{pmatrix},\ \ \ \ 
\langle \phi'_{\rm sol} \rangle =\varphi'_{\rm sol}
\begin{pmatrix}
 1 \\
 3\\
 -1
\end{pmatrix}.
\label{CSD3}
\ee
An important point we would like to emphasise is that, as discussed in
Appendix~\ref{app:CGs}, in the $S_4$ basis employed in this paper
the above vacuum alignments preserve the generator product $SU$,
i.e. $SU \langle \phi'_{\rm atm} \rangle = \langle \phi'_{\rm atm} \rangle $
and  $SU \langle \phi'_{\rm sol} \rangle = \langle \phi'_{\rm sol} \rangle $,
but break $T$ and $U$ separately. Assuming that the charged lepton mass matrix
is diagonal, the preserved $S_4$ subgroup $SU$ is instrumental
in enforcing TM$_1$ mixing as in the semi-direct model of~\cite{Luhn:2013vna}.
However, unlike  \cite{Luhn:2013vna}, this model involves two right-handed
neutrinos which are assumed to have a diagonal mass matrix $M_R$.

The $S_4$ singlet contraction 
${\bf 3}^{{\prime}} \otimes {\bf 3}^{{\prime}} 
~\rightarrow ~{\bf   1}^{\phantom{\prime}}$  
implies $(L\phi')_{\bf 1}  = L_1\phi'_1 + L_2\phi'_3 +L_3\phi'_2 $  
(see Appendix~\ref{app:CGs}), 
which leads to the Dirac neutrino mass matrix~$m^D$, together with a diagonal
right-handed neutrino mass matrix~$M_R$,
\begin{equation}
	m^D = \pmatr{0 & b \\ -a & -b \\ a & 3b } \equiv \pmatr{0 & b \\ a & b \\ a & 3b } ,
	\qquad
	M_{R}=
\left( \begin{array}{cc}
M_{\rm atm} & 0 \\
0 & M_{\rm sol}
\end{array}
\right),
			\label{mDn0}
\end{equation}
where the equivalence above follows after multiplying $L_2$ by a minus sign.
The seesaw mechanism $m^{\nu}=-m^DM_{R}^{-1}{m^D}^T$ implies\footnote{We
  follow the Majorana mass convention $- \frac{1}{2}\overline{\nu_L} m^{\nu}
  \nu_{L}^c $.} 
\begin{equation}
	m^\nu = m_a 
	\left(
\begin{array}{ccc}
	0&0&0\\0&1&1\\0&1&1 
	\end{array}
\right)
	+ m_b e^{i\eta } 
	\left(
\begin{array}{ccc}
	1&1&3\\1&1&3\\3&3&9
	\end{array}
\right),
	\label{eq:mnu2p10}
\end{equation}
where without loss of generality, $m_a=|a|^2/M_{\rm atm}$, $m_b=|b|^2/M_{\rm
sol}$ may be taken to be real and positive and $\eta$ is a real phase parameter.
Eq.~\eqref{eq:mnu2p10} with $\eta = -2\pi/3$ gives a phenomenologically
successful and predictive description of neutrino masses and lepton mixing
parameters, as first discussed in~\cite{King:2013iva}. In fact the neutrino
mass matrix in Eq.~\eqref{eq:mnu2p10} with $ \eta = \pm 2\pi/3 $ is one of the
two CSD3 forms first discussed in~\cite{King:2013iva}.

The main point we wish to emphasise is that the neutrino mass matrix in
Eq.~\eqref{eq:mnu2p10}, which is related to that in Eq.~\eqref{eq:mnu2p100} by
the permutation $L_2\leftrightarrow L_3$, leads to phenomenologically
successful predictions for neutrino parameters for a phase $\eta = \pm 2\pi/3$. 
In Table~\ref{tab:model} we compare predictions from the two forms of CSD3
neutrino mass matrix in Eq.~\eqref{eq:mnu2p10} and Eq.~\eqref{eq:mnu2p100} for
some benchmark input parameters $m_a, m_b, \eta$. The two types of CSD3 yield
identical predictions for the reactor and solar angles as well as the neutrino
masses, for the same values of $m_a, m_b$, while the predictions for the
atmospheric angle have the same values of $\sin 2\theta_{23}$ but are in
different octants of $\theta_{23}$. It is clear that both types of CSD3 give
good predictions for lepton mixing angles, assuming that $\eta = \pm 2\pi/3$.
In both examples in Table~\ref{tab:model} the CP phase is predicted to be
$\delta_{\mathrm{CP}}\approx -\pi/2$.\footnote{In addition, the CSD3 in
  Eq.~\eqref{eq:mnu2p10} predicts the Majorana phase $\beta = -71.9^{\circ}$
  (as compared to $\beta = 71.9^{\circ}$ with Eq.~\eqref{eq:mnu2p100}) which
  is not shown in the Table since the neutrinoless double beta decay parameter
  is $m_{ee}=m_b= 2.684$ meV for the above parameter set which is practically
  impossible to measure in the foreseeable future.}
For the original CSD3, $\eta = 2\pi/3$ is identified as the leptogenesis
phase and the baryon asymmetry of the universe leads to a determination of the
lighter atmospheric neutrino mass $M_{\rm atm}=4\times 10^{10}$
GeV~\cite{Bjorkeroth:2015tsa}. For the new type of CSD3 here we expect
leptogenesis to fix the lighter solar right-handed neutrino mass to be $M_{\rm
  sol}=4\times 10^{10}$ GeV due to the preferred opposite value of the
leptogenesis phase $\eta = -2\pi/3$.

\begin{table}[ht]
\renewcommand{\arraystretch}{1.2}
\centering
\footnotesize
\begin{tabular}{| c c c | c c c c c c c |}
\hline
\rule{0pt}{4ex}%
\makecell{$m_a$ \\ {\scriptsize (meV)}} & \makecell{$m_b$ \\ {\scriptsize (meV)}} & 
\makecell{$\eta$  \\ {\scriptsize (rad)}}  	& \makecell{$\theta_{12}$ \\ {\scriptsize ($^{\circ}$)}} & \makecell{$\theta_{13}$ \\ {\scriptsize ($^{\circ}$)}}  & \makecell{$\theta_{23}$ \\ {\scriptsize ($^{\circ}$)}} & \makecell{$\delta_{\mathrm{CP}}$ \\ {\scriptsize ($^{\circ}$)}} & \makecell{$m_1$ \\ {\scriptsize (meV)}}
& \makecell{$m_2$ \\ {\scriptsize (meV)}} & \makecell{$m_3$ \\ {\scriptsize (meV)}} \\ [2ex] \hline 
\rule{0pt}{4ex}%
26.57		& 2.684		& $ -\dfrac{2\pi}{3} $	& 34.3		& 8.67		& 44.2		& -93.3		& 0 & 8.59		& 49.8 \\[1.7ex]
\hline
26.57		& 2.684		& $ \dfrac{2\pi}{3} $	& 34.3		& 8.67		& 45.8		& -86.7		& 0 & 8.59		& 49.8 \\[1.7ex]
\hline
Value & from & \cite{Gonzalez-Garcia:2014bfa} & 33.48$^{+0.78}_{-0.75}$ & 8.50$^{+0.20}_{-0.21}$ &
42.3$^{+3.0}_{-1.6}$  &  -54$^{+39}_{-70}$ & 0 & 8.66$\pm 0.10$ & 49.57$\pm 0.47$ \\
\hline
\end{tabular}
\caption{Benchmark parameters and predictions for CSD3 in
  Eq.~\eqref{eq:mnu2p10} used in this paper   (second line) with a fixed phase
  $ \eta = -2\pi/3 $, as compared to the version of CSD3 in
  Eq.~\eqref{eq:mnu2p100} (third line) with a fixed phase of $ \eta = 2\pi/3$. 
These predictions, which depend on the theoretical fit~\cite{Bjorkeroth:2014vha},
as well as possible charged lepton and renormalisation group
corrections~\cite{Boudjemaa:2008jf}, may be compared to the global best fit values
from~\cite{Gonzalez-Garcia:2014bfa} (for $m_1=0$), given in the fourth line
(see also~\cite{Capozzi:2013csa,Forero:2014bxa}).}  
\label{tab:model}
\end{table}

In Appendix~\ref{analytic}, the mass matrix in Eq.~\eqref{eq:mnu2p10} is
generalised to a new type of CSD$n$, and analytic formulas for neutrino
masses and lepton mixing parameters are presented for any real value of $n$
(although we are only interested in $n=3$ here). The results may be compared
to the numerical results in~\cite{Bjorkeroth:2014vha} and the analytic
formulas in~\cite{King:2015dvf} for the original version of CSD$n$ based on a
generalisation of Eq.~\eqref{eq:mnu2p100}.

%%%%%%%%%%%%%%%%%%%%%%%%%%%%%%%%%%%%%%%%%%%%%

%%%%%%%%%%%%%%%%%%%%%%%%%%%%%%%%%%%%%%%%%%%%%

%%%%%%%%%%%%%%%%%%%%%%%%%%%%%%%%%%%%%%%%%%%%%

\section{\label{sec:align}Vacuum alignment for CSD3}

In our setup, we rely on the supersymmetric $F$-term alignment mechanism to
generate the 
appropriate symmetry breaking flavon VEVs. The required driving fields are
denoted by $X_i$, $Y_i$, $Z_i$, where the subscript $i$ indicates its $S_4$
representation. We derive all necessary alignments in a short sequence of
steps. Commencing with the primary alignments of triplets flavons, we proceed to
generate alignments of doublet flavons. In a final step,  the $SU$ preserving
CSD3 alignments are obtained from $SU$ symmetric $F$-term conditions. 
Our notation is such that the three primary triplet flavons are denoted by 
$\phi'_{S,U}\sim{\bf{3'}}$, $\phi_T \sim {\bf{3}}$ and 
$\phi'_t\sim {\bf{3'}}$. The doublet flavons, which are obtained from the
primary ones, are  $\rho^{}_{S,U}\sim{\bf{2}}$ and $\rho^{}_{t}\sim{\bf{2}}$. Here,
the indices ($S,U,T$ and $t$) show the symmetry preserving generators, where
$t$ corresponds to $T$ multiplied by a $Z_3$ generator which is not part of
$S_4$. In addition to the triplet and doublet flavons, we also introduce the $S_4$
singlet flavons $\xi_T \sim {\bf{1}}$ and $\xi_{S,U} \sim  {\bf{1}}$.

The primary triplet alignments are derived from simply coupling the square of
a flavon triplet to a single driving field $X_i$. The resulting $F$-term conditions
depend on the $S_4$ representation of $X_i$, and the most general solutions of
these conditions are given as follows.
\bea
X_{3'} (\phi'_{S,U})^2 & ~~~~~\longrightarrow~~~~~ & 
\begin{pmatrix} 1\\\omega^n\\\omega^{2n}\end{pmatrix}\ ,\\
X_2 (\phi_T)^2 & ~~~~~\longrightarrow ~~~~~ & \begin{pmatrix} 1\\0\\0\end{pmatrix},
\begin{pmatrix} 1\\-2\omega^n \\-2\omega^{2n}\end{pmatrix} \ ,\\
X_1 (\phi'_{t})^2 & ~~~~~\longrightarrow~~~~~ & 
\begin{pmatrix} 0\\0\\1\end{pmatrix},
\begin{pmatrix} 0\\1\\0\end{pmatrix},
\begin{pmatrix} 2\\2x\\-1/x\end{pmatrix}\ ,
\eea
where the alignments are only fixed up to an integer ($n\in \mathbb Z$) or
continuous ($x\in \mathbb R$) parameter, with $\omega\equiv e^{2\pi i/3}$.

We emphasise that all solutions of the $\phi_T$ alignments are related by
$S_4$ transformations. It is therefore possible to choose the direction
$\langle \phi_T\rangle \propto (1,0,0)^T$ without loss of
generality. Moreover, the alignments of $\phi'_{S,U}$ can be brought to the
standard $(1,1,1)^T$ form by a $T$ transformation which does not affect
the $\phi_T$ alignment. 
Finally, the so-selected alignments of $\phi_T$ and $\phi'_{S,U}$ do not
change their form (up to a possible overall sign) under application of a $U$
transformation. This fact allows us to get rid of the ambiguity of the
$\phi'_t$ alignment: the third alignment $(2,2x,-1/x)^T$ can be removed by
requiring orthogonality with $\langle\phi_T\rangle$, which can be enforced in
a straightforward way by the term 
\be X_{1'} \phi_T   \phi'_t\ ,
\ee
in the driving potential. 
Then, a $U$ transformation can be applied to choose the alignment
$\langle \phi'_t \rangle\propto (0,1,0)^T$ without loss of generality.  
We can thus make use of the following three primary alignments 
\begin{equation}\label{eq:primary}
\langle \phi'_{S,U} \rangle  ~=~ \varphi'_{S,U} \begin{pmatrix} 1\\1\\1\end{pmatrix} ,
\qquad
\langle \phi_T \rangle  ~=~\varphi_T  \begin{pmatrix} 1\\0\\0\end{pmatrix} ,
\qquad
\langle \phi'_t \rangle  ~=~\varphi'_t \begin{pmatrix} 0\\1\\0\end{pmatrix} ,
\end{equation}
to generate new alignments, which together with the primary ones can be used
in constructing our CSD3 model of leptons. 
Note that $\langle \phi'_{S,U} \rangle $ preserves $S,U$ while 
$\langle \phi_T \rangle$ preserves $T$ as discussed in Appendix~\ref{app:CGs}.

The secondary alignments of the doublet flavons $\rho^{}_{S,U}$ and
$\rho^{}_{t}$ originate in the driving terms
\bea\label{eq:ali_rho}
Y_{3} \phi'_{S,U} \rho^{}_{S,U} \ , \qquad
Y_{3'} \left(\xi^{}_T \phi'_t   - \phi_T  \rho^{}_t  \right) \ ,
\eea
where $\xi_T$ represents an $S_4$ singlet flavon which does not affect the
alignment of~$\rho_t$. We remark that all dimensionless coupling
constants of the flavon potential are suppressed for the sake of notational
clarity. It is, however, important to keep in mind that such couplings are
real in our setup with imposed CP symmetry. 
A straightforward calculation shows that the $F$-term conditions resulting
from Eq.~\eqref{eq:ali_rho} determine the doublet alignments uniquely to
\begin{equation}\label{eq:11}
\langle \rho^{}_{S,U} \rangle ~=~ \varrho^{}_{S,U} \begin{pmatrix} 1\\1\end{pmatrix} ,
\qquad
\langle \rho^{}_t \rangle ~=~ \varrho^{}_t  \begin{pmatrix} 0\\1\end{pmatrix} .
\end{equation}
We point out that the doublet flavon $\rho^{}_t$ is actually not required in
constructing the CSD3 alignments. However, it can be used in the charged lepton
sector to generate the muon and electron masses.\footnote{For the tau mass
we can use the triplet flavon $\phi'_t$. The product $\phi'_t \rho^{}_t$ yields and
effective alignment in the $(0,0,1)^T$ direction and can be used to generate
the muon mass. Finally the product $\phi'_t \rho^{}_t \rho^{}_t$ gives rise to an
effective vacuum alignment in the $(1,0,0)^T$ direction so that it can be
adopted to give mass to the electron. (In principle we could also use the 
$\phi_T$ flavon for the electron, but the relative suppression of the 
electron mass with respect to the tau and muon mass would require an unnatural
hierarchy between the VEVs of $\phi_T$ and $\phi'_t$.)}

Turning to the derivation of the CSD3 alignments, we first consider the
contraction of $\phi'_{S,U}$ and $\phi_T$ to a ${\bf{3'}}$ of $S_4$,
\begin{equation}
\left[\langle \phi'_{S,U} \rangle \cdot \langle \phi_T \rangle \right]_{3'} 
~~ \propto~~
 \begin{pmatrix}0\\1\\-1 \end{pmatrix} .
\end{equation}
Although the flavon direction $\langle \phi_T \rangle$ does not respect the
$SU$ symmetry, its product with $\langle \phi'_{S,U} \rangle$, contracted to a
${\bf{3'}}$, yields an $SU$ invariant direction. From this result, we
immediately see that the driving term
\begin{eqnarray}\label{eq:ali_atm}
Z_{3'} \left( \phi'_{S,U} \phi^{}_T- \xi^{}_{S,U} \phi'_{\mathrm{atm}} \right) \ ,
\end{eqnarray}
with $\xi^{}_{S,U}$ being an $S_4$ singlet flavon field, generates the alignment 
\begin{equation}\label{eq:secondary}
\langle \phi'_{\mathrm{atm}} \rangle ~=~\varphi'_{\mathrm{atm}} \begin{pmatrix} 0\\1\\-1\end{pmatrix} .
\end{equation}
Similarly, we can consider the product of
$\phi'_{\mathrm{atm}} $ and $\phi'_t$ to a ${\bf{3'}}$ of $S_4$, 
\begin{equation}
\left[\langle \phi'_{\mathrm{atm}} \rangle\cdot \langle \phi'_t \rangle
  \right]_{3'} ~~ \propto~~ 
 \begin{pmatrix}1\\0\\2 \end{pmatrix} .\label{eq:102}
\end{equation}
Again, the flavon direction $\langle \phi'_t \rangle$ does not respect $SU$, yet its
product with $\langle\phi'_{\mathrm{atm}}\rangle$ to a ${\bf{3'}}$ does. 
In order to realise the CSD3 alignment $\phi'_{\mathrm{sol}}$, we use the
particular $SU$ preserving product of Eq.~\eqref{eq:102} as well as the
doublet flavon $\rho^{}_{S,U}$ (whose VEV is invariant under $SU$) in the
driving term 
\be
\tilde Z_{3'} \left(   \phi'_{\mathrm{atm}} \phi'_t- 
 \rho^{}_{S,U}  \,\phi'_{\mathrm{sol}}  \right)
 \ . \label{eq:csd3}
\ee
To see this, we insert the already aligned flavon directions into
Eq.~\eqref{eq:csd3}. This gives the $F$-term conditions for $\langle 
\phi'_{\mathrm{sol}} \rangle = (\beta_1,\beta_2,\beta_3)^T$ 
\be
\varphi'_{\mathrm{atm}}\,\varphi'_t \begin{pmatrix}
1\\0\\2
\end{pmatrix} 
-  \varrho^{}_{S,U}
\begin{pmatrix} \beta_2 + \beta_3 \\  \beta_3 + \beta_1 \\  \beta_1 + \beta_2
\end{pmatrix}   ~=~
\begin{pmatrix}
0\\0\\0
\end{pmatrix},
\ee 
which uniquely specify the alignment to the CSD3 one
\be
\langle \phi'_{\mathrm{sol}} \rangle  ~=~ \varphi'_{\mathrm{sol}} \begin{pmatrix} 1\\3\\-1\end{pmatrix} .
\ee
Furthermore, the VEVs 
$\varphi'_{\mathrm{atm}}$ and $\varphi'_{\mathrm{sol}}$ (including the phase) 
are related via
\be
{\varphi'_{\mathrm{sol}}} ~=~
\frac{\varphi'_t}{2\,\varrho^{}_{S,U}}~{\varphi'_{\mathrm{atm}}}\ .
\ee

Having completed the discussion of the vacuum alignment for supersymmetric
CSD3 models, we conclude this section by collecting all terms of the flavon
sector in the flavon superpotential. Suppressing all coupling coefficients
(which are real in the case of a CP symmetric setup), we have the superpotential
\bea
W_{0}^{\mathrm{flavon}}&\sim& 
X_{3'} (\phi'_{S,U})^2 + X_2 (\phi^{}_T)^2 + X_1 (\phi'_t)^2 +
X_{1'}\phi^{}_T\phi'_t \nonumber \\
&& + Y_3 \phi'_{S,U} \rho^{}_{S,U} + Y_{3'} (\xi^{}_T \phi'_t - \phi^{}_T
\rho^{}_t) \label{eq:flavonpot0} \\
&&+ Z_{3'} (\phi'_{S,U}\phi^{}_T - \xi^{}_{S,U} \phi'_{\mathrm{atm}}) 
+ \tilde Z_{3'} (\phi'_{\mathrm{atm}} \phi'_t - \rho^{}_{S,U}
\phi'_{\mathrm{sol}}) \ . \nonumber
\eea
It is important to notice that the flavon potential of
Eq.~\eqref{eq:flavonpot0} contains only renormalisable terms. As a
consequence, the CSD3 alignments derived from the corresponding $F$-term
conditions should be relatively robust when implemented into a concrete model.

%%%%%%%%%%%%%%%%%%%%%%%%%%%%%%%%%

%%%%%%%%%%%%%%%%%%%%%%%%%%%%%%%%%

%%%%%%%%%%%%%%%%%%%%%%%%%%%%%%%%%

\section{A concrete model of CSD3}\label{concrete}

In order to define a model, it is necessary to specify its particle content as
well as all symmetries which constrain the couplings of the fields. In
Eq.~\eqref{eq:flavonpot0}, we have already stated the flavon superpotential
for generating the CSD3 alignments. By construction, these terms are symmetric 
under the imposed $S_4$ family symmetry. Furthermore, it is possible to
introduce a $U(1)$ symmetry which allows for all terms of
Eq.~\eqref{eq:flavonpot0}. Such a $U(1)$ must, however, also be consistent
with the superpotential terms of the lepton sector. Following the discussion of
the Littlest Seesaw model~\cite{King:2015dvf}, we demand the superpotential terms
\bea\label{eq:Wlepton1}
W^{\mathrm{lepton}}_0 &=& ~
\frac{y'_\tau}{\Lambda} L H_d E^c_3 \, \phi'_t + 
\frac{y'_\mu}{\Lambda^2} L H_d E^c_2 \, \phi'_t \, \rho^{}_t + 
\frac{y'_e}{\Lambda^3} L H_d E^c_1 \, \phi'_t \, (\rho^{}_t)^2 \\\nonumber
&&\!\!\!+ \frac{y_{\mathrm{atm}}}{\Lambda} L H_u N^c_{\mathrm{atm}} \, \phi'_{\mathrm{atm}} 
+ \frac{y_{\mathrm{sol}}}{\Lambda} L H_u N^c_{\mathrm{sol}} \, \phi'_{\mathrm{sol}} 
+ \xi^{}_{\mathrm{atm}} \, N^c_{\mathrm{atm}} N^c_{\mathrm{atm}} + \xi^{}_{\mathrm{sol}}\, N^c_{\mathrm{sol}} N^c_{\mathrm{sol}} \ .
\eea
Here we assume the Higgs doublets $H_{u}$ and $H_{d}$ to transform trivially
under $S_4$ as well as any additional $U(1)$ symmetry. 
The neutrino sector of Eq.~\eqref{eq:Wlepton1} contains the typical CSD3 Dirac
mass terms, while the Majorana mass terms arise from the VEVs of the $S_4$
singlet flavons $\xi_{\mathrm{atm}}$ and
$\xi_{\mathrm{sol}}$.  We suppress the 
dimensionless Yukawa couplings in the Majorana sector for
brevity.\footnote{Replacing the singlet flavon (Majoron) fields 
$\xi$ by bare mass parameters~$M$, it is possible to show that the flavon
  superpotential would include additional renormalisable terms which spoil our
  successful method of generating the CSD3 alignment.} Considering the
charged lepton sector, we choose the right-handed electrons as $S_4$
singlets, while the three generations of left-handed lepton doublets $L_i$ are
combined into the $S_4$ triplet ${\bf{3'}}$. Contracting $L$ with $\langle
\phi'_t\rangle$ to an $S_4$ invariant projects out the third family
$L_3$. Similarly, the $S_4$  products 
$L \langle \phi'_t\rangle \langle \rho_t\rangle $ and 
$L \langle \phi'_t\rangle \langle \rho_t\rangle^2 $ project out $L_2$ and
$L_1$, respectively. We thus obtain a diagonal charged lepton mass matrix in
which the hierarchy of masses results from different powers of the suppression 
factor~$1/\Lambda$. 

The operators of Eq.~\eqref{eq:Wlepton1} put further constraints on a possible
$U(1)$ symmetry. Counting the number of fields and comparing this to the
number of constraints from Eqs.~(\ref{eq:flavonpot0},\ref{eq:Wlepton1}), we
can determine the maximal $U(1)$ symmetry which is allowed in our setup. With
25~fields and 18~independent terms, we obtain 7 free parameters which
specify the most general $U(1)$ symmetry. Expressing its charges in terms of
the parameters $x_{1,2}$ and $z_{1,2,3,4,5}$, we list the complete charge
assignments in Table~\ref{tab:charges}. 

\begin{table}[t!]
\begin{center}
\begin{tabular}{|c|c|c|c||c|c|c|c|c|c|} \hline
\multicolumn{2}{|c|}{fields}&  $S_4$ & $U(1)_{\mathrm{}}$ 
& $\!U(1)_{x}\!$ & $\!Z_3^{(1)}\!$ & $\!Z_3^{(2)}\!$ & $\!Z_3^{(3)}\!$ &
$\!Z_3^{(4)}\!$ & $\!Z_3^{(5)}\!$ \\\hline
\multirow{8}{*}{\rotatebox[origin=c]{90}{Higgs \& leptons}}
& $L$ & ${\bf{3'}}$  &$-x_1+z_1$    
&  $1$ &  $1$ &  $0$ & $0$ &  $0$ &  $0$  \\
& $E^c_3$ & ${\bf{1}}$  &$x_1-z_1-z_3$    
&  $-1$ &  $2$ &  $0$ & $2$ &  $0$ &  $0$  \\
& $E^c_2$ & ${\bf{1}}$  &$x_1-x_2-z_1-2z_3-z_4+z_5$    
&  $-4$ &  $2$ & $0$ & $1$ &  $2$ &  $1$  \\
& $E^c_1$ & ${\bf{1}}$  &$x_1-2x_2-z_1-3z_3-2z_4+2z_5$    
&  $-7$ &  $2$ & $0$ & $0$ &  $1$ &  $2$  \\
& $N^c_{\mathrm{atm}}$ & ${\bf{1}}$  &$-z_1$    
&  $0$ &  $2$ & $0$ & $0$ &  $0$ &  $0$  \\
& $N^c_{\mathrm{sol}}$ & ${\bf{1}}$  &$-z_2$    
&  $0$ &  $0$ & $2$ & $0$ &  $0$ &  $0$  \\
& $H_d$ & ${\bf{1}}$  &$0$    
&  $0$ &  $0$ &  $0$ & $0$ &  $0$ &  $0$  \\
& $H_u$ & ${\bf{1}}$  &$0$    
&  $0$ &  $0$ &  $0$ & $0$ &  $0$ &  $0$  \\\hline
\multirow{11}{*}{\rotatebox[origin=c]{90}{flavon fields}}
& $\phi'_{S,U}$ & ${\bf{3'}}$ & $x_1+x_2$ 
& $2$ & $0$ &$0$ &$0$ &$0$ &$0$ \\
& $\rho^{}_{S,U}$ & ${\bf{2}}$ & $z_1-z_2+z_3$ 
& $0$ & $1$ &$2$ &$1$ &$0$ &$0$ \\
& $\xi^{}_{S,U}$ & ${\bf{1}}$ & $z_5$ 
& $0$ &  $0$ &$0$ &$0$ &$0$ &$1$ \\
& $\phi_T$ & ${\bf{3}}$ & $-x_2+z_5$ 
&  $-3$ &$0$ &$0$ &$0$ &$0$ &$1$ \\
& $\xi^{}_T$ & ${\bf{1}}$ & $z_4$ 
& $0$ & $0$ & $0$ & $0$ & $1$ & $0$  \\\cline{2-10}
& $\phi'_t$ & ${\bf{3'}}$ & $z_3$ 
& $0$ & $0$ & $0$ & $1$ & $0$ & $0$  \\
& $\rho^{}_t$ & ${\bf{2}}$ & $x_2+z_3+z_4-z_5$ 
& $3$ & $0$ & $0$& $1$& $1$& $2$ \\
& $\phi'_{\mathrm{atm}}$ & ${\bf{3'}}$ &  $x_1$ 
& $-1$ &$0$ &$0$ &$0$ &$0$ &$0$ \\
& $\phi'_{\mathrm{sol}}$  & ${\bf{3'}}$ & $x_1-z_1+z_2$  
& $-1$ & $2$  & $1$& $0$& $0$& $0$\\
& $\xi_{\mathrm{atm}}$ & ${\bf{1}}$ &  $2z_1$ 
& $0$ & $2$ & $0$ & $0$ & $0$ & $0$  \\
& $\xi_{\mathrm{sol}}$  & ${\bf{1}}$ & $2z_2$  
& $0$ & $0$ & $2$ & $0$ & $0$ &
$0$   \\\hline
\multirow{9}{*}{\rotatebox[origin=c]{90}{driving fields}}
& $X_{3'}$ & ${\bf{3'}}$  &$-2x_1-2x_2$    
&  $-4$ &  $0$ &  $0$ & $0$ &  $0$ &  $0$  \\
& $X_{2}$ & ${\bf{2}}$  &$2x_2-2z_5$    
&  $6$ &  $0$ &  $0$ & $0$ &  $0$ &  $1$  \\
& $X_{1}$ & ${\bf{1}}$  &$-2z_3$    
&  $0$ &  $0$ &  $0$ &$1$ &  $0$ &  $0$  \\
& $X_{1'}$ & ${\bf{1'}}$  &$x_2-z_3-z_5$    
&  $3$ &  $0$ &  $0$ &$2$ &  $0$ &  $2$  \\
& $Y_{3}$ & ${\bf{3}}$  &$-x_1-x_2-z_1+z_2-z_3$    
&  $-2$ &  $2$ &  $1$ & $2$ &  $0$ &  $0$  \\
& $Y_{3'}$ & ${\bf{3'}}$  &$-z_3-z_4$    
&  $0$ &  $0$ &  $0$ & $2$ &  $2$ &  $0$  \\
& $Z_{3'}$ & ${\bf{3'}}$  &$-x_1-z_5$    
&  $1$ &  $0$ &  $0$ & $0$ &  $0$ &  $2$  \\
& $\tilde Z_{3'}$ & ${\bf{3'}}$  &$-x_1-z_3$    
&  $1$ &  $0$ &  $0$ & $2$ &  $0$ &  $0$  \\
& $X_0$ & ${\bf{1}}$  &$0$    
&  $0$ &  $0$ &  $0$ & $0$ &  $0$ &  $0$  \\\hline
\end{tabular}
\end{center}
\caption{\label{tab:charges}The particle content and symmetries of our CSD3
  model. $U(1)_{\mathrm{}}$ denotes the most general symmetry
  consistent with the terms of
  Eqs.~(\ref{eq:flavonpot0},\ref{eq:Wlepton1}). $U(1)_x$ is specified by 
  setting $x_1=-1$, $x_2=3$ and $z_i=0$. The $Z_3^{(i)}$ symmetries
  are $Z_3$ subgroups of $U(1)_{\mathrm{}}$ with all parameters set to zero
  except for $z_i=1$. In addition, we assume a standard $U(1)_R$ symmetry
  with the charge assignments: $+1$  for lepton, $+2$ for driving fields,
  $0$ for Higgs and flavon fields.}
\end{table}

Imposing this general $U(1)_{\mathrm{}}$ for arbitrary parameters $x_{1,2}$ and
$z_{1,2,3,4,5}$ is tantamount to imposing seven independent $U(1)$
symmetries. It is straightforward to show that such a powerful symmetry, while
being consistent with all term of
Eqs.~(\ref{eq:flavonpot0},\ref{eq:Wlepton1}), does not allow for any other
relevant term. We have checked this result explicitly for terms with up to
five flavon fields, finding no extra term at all.\footnote{Imposing only one
  particular $U(1)$ symmetry rather than 
  seven independent $U(1)$s, it is also possible to forbid all relevant
  unwanted operators. For instance, with the somewhat arbitrary choice 
  $(x_1,x_2,z_1,z_2,z_3,z_4,z_5)=(4,16,-61,88,53,-61,7)$, the driving potential
does not have any non-renormalisable operator with three flavon
fields. Likewise, the first new Dirac-type terms of the lepton sector involve
four flavons and are therefore highly suppressed. For the right-handed neutrinos, we
encounter a new contribution to $N^c_{\mathrm{atm}}N^c_{\mathrm{atm}}$
with two flavons, however, the first off-diagonal term
$N^c_{\mathrm{atm}}N^c_{\mathrm{sol}}$ already requires five flavons.}

As discussed in~\cite{King:2015dvf}, the Littlest Seesaw requires the 
relative phase factor~$\omega=e^{2\pi i/3}$ between the two contributions to the
effective light neutrino mass matrix. In a CP conserving setup, such a phase
factor can only originate in complex flavon VEVs. In order to predict phases,
it is necessary to find a way of driving flavon VEVs to certain values with
given phases. An obvious option is to introduce a completely neutral driving
field $X_0$ which couples to both, some power of a flavon field $\phi$ as well as a
bare mass parameter. For instance, 
$X_0 (\phi^2 - M^2)$ entails a real VEV for the flavon
$\phi$ provided that $M$ is real. Such a method has been applied previously,
e.g. in~\cite{Antusch:2011sx}. In order to drive a flavon VEV to a complex
value whose phase factor is~$\omega^k$, it is suggestive to make use of
couplings such as $X_0 (\phi^3/\Lambda - M^2)$, see
e.g.~\cite{Luhn:2013vna}. Clearly, this structure is forbidden if the flavon
$\phi$ carries a non-trivial $U(1)$ charge. However, a non-trivial $Z_3$
charge is possible; in fact, it is even necessary in order to forbid the quadratic term
$X_0\phi^2$. 

On the right-hand side of Table~\ref{tab:charges}, we have defined particular
subgroups of the general $U(1)_{\mathrm{}}$ which, as mentioned earlier, can be
understood as seven independent $U(1)$ symmetries. The $U(1)_x$ symmetry is
defined by choosing $x_1=-1$, $x_2=3$ and $z_{i}=0$. The $Z_3^{(i)}$
symmetries are obtained as discrete subgroups of the general
$U(1)_{\mathrm{}}$ with all 
parameters set to zero except for $z_i=1$. Imposing only $U(1)_x$ and the five
$Z_3^{(i)}$ symmetries, it is possible to drive the VEVs of the flavons with
zero $U(1)_x$ charge to values with a phase factor~$\omega^k$. As the
so-reduced symmetry could, in principle, allow for other new terms in the 
superpotential, we have to check for such unwanted operators. In addition to
the terms of Eq.~\eqref{eq:flavonpot0}, we find the following cubic terms in the flavon
potential,
\be
W^{\mathrm{flavon}}_1\sim X_0 \left[ 
\frac{(\xi_{\mathrm{atm}})^3
+(\xi_{\mathrm{sol}})^3
+(\xi^{}_T)^3
+(\xi^{}_{S,U})^3
+(\phi'_t)^3
+(\rho^{}_{S,U})^3
+\phi'_{S,U} (\phi'_{\mathrm{atm}})^2
}{\Lambda}
-M^2 \right].\label{eq:drive}
\ee
All terms with one driving field coupling to four flavons are forbidden, while
there exist many allowed, though strongly suppressed, terms with five
flavons. In the Dirac-type terms of the lepton sector, the first new terms
involve four flavon fields and are therefore highly suppressed. Finally, we
find extra contributions to the mass terms of the right-handed neutrinos with four
or more flavons. The complete model based on the $U(1)_x \times Z_3^{(1)}
\times Z_3^{(2)} \times Z_3^{(3)} \times Z_3^{(4)} \times Z_3^{(5)} $ symmetry
is therefore given by the superpotentials
\bea
W^{\mathrm{flavon}}_{} &=& W^{\mathrm{flavon}}_{0} \, +~ W^{\mathrm{flavon}}_{1}
\,+~ \left(\frac{1}{\Lambda^3} \,X\, \phi^5 ~+~ \cdots \right) \ ,\\
W^{\mathrm{lepton}}_{} &=& W^{\mathrm{lepton}}_{0} \, + ~
\left( \frac{1}{\Lambda^4} \, LH_d E^c_i \, \phi^4 ~+~ 
\frac{1}{\Lambda^4} \, LH_u N^c_i \, \phi^4 ~+~ 
\frac{1}{\Lambda^3} \, N^c_i N^c_j \, \phi^4 ~+~ \cdots \right) \ ,~~~~~
\eea
where the higher order terms in brackets are only written schematically with $X$ or $\phi$
representing any of the driving or flavon fields of the model.

These observations show that the reduced symmetry on the
right-hand side of Table~\ref{tab:charges} is sufficient to control the
coupling of driving, flavon and lepton fields. Moreover, Eq.~\eqref{eq:drive}
allows us to constrain the VEVs of the flavons. 
The existence of the mixed term $\phi'_{S,U} (\phi'_{\mathrm{atm}})^2$ in
Eq.~\eqref{eq:drive} follows from the particular charge assignment under the $U(1)_x$
symmetry. However, inserting the vacuum alignment, this term vanishes
identically. Hence we can ignore it in the following. 
Adding six copies of the driving field $X_0$, we obtain six independent
$F$-term equations which decouple if linearly combined. 
Then, the VEVs of the flavons $\xi_{\mathrm{atm}}$,
$\xi_{\mathrm{sol}}$, $\xi^{}_T$, $\xi^{}_{S,U}$, $\phi'_t$ and
$\rho^{}_{S,U}$ are driven to values where the phase factor is some power of
$\omega$.\footnote{Notice that -- with the respective alignments --
$\langle\phi'_t\rangle^3$ as well as  $\langle\rho^{}_{S,U}\rangle^3$
have non-vanishing contractions to an $S_4$ singlet. This is not the case for
$\langle \phi'_{S,U}\rangle^3$.}  
Due to the symmetries, many of these phase factors can however be removed. For
instance, if $\xi_{\mathrm{atm}}$ has a phase factor~$\omega^k$, this can
be modified to~$\omega^0$ by a $Z_3^{(1)}$ transformation. Since
$\xi_{\mathrm{atm}}$ is uncharged under any 
of the other symmetries, a $Z_3^{(2)}$ transformation can be applied without
modifying the trivial phase of  $\xi_{\mathrm{atm}}$. On the other hand, such
a $Z_3^{(2)}$ transformation  can remove the phase of
$\xi_{\mathrm{sol}}$. This procedure can be applied further to render real the VEVs of
$\xi_{\mathrm{atm}}$, $\xi_{\mathrm{sol}}$, $\xi^{}_T$, $\xi^{}_{S,U}$ as well
as $\phi'_t$. Having exhausted all $Z_3^{(i)}$ symmetries, the phase factor
of $\rho^{}_{S,U}$ cannot be removed. 
Similarly to the $Z_3^{(i)}$ transformations, the $U(1)_x$ symmetry
can be used to remove the phase of $\langle\phi'_{\mathrm{atm}}\rangle$. 
Defining the phase factor of $\langle\phi'_{S,U}\rangle$ to be $e^{i\alpha}$,
the phases of the VEVs of the remaining flavons $\phi^{}_T$, $\rho^{}_t$ and
$\phi'_{\mathrm{sol}}$ are fixed by
Eqs.~(\ref{eq:ali_atm},\ref{eq:ali_rho},\ref{eq:csd3}), respectively. 
In summary, we can work in a basis where the VEVs of all 
flavons are real except for the following list
\be
\frac{\varrho^{}_{S,U} }{|\varrho^{}_{S,U}|} ~=~ \omega^k \ ,\qquad
\frac{\varphi'_{\mathrm{sol}}}{|\varphi'_{\mathrm{sol}}|} ~=~ \omega^{-k} \ ,
\ee
\be
\frac{\varphi'_{S,U} }{|\varphi'_{S,U}|} ~=~ \frac{\varrho^{}_t }{|\varrho^{}_t|} ~=~ e^{i\alpha} \ ,\qquad
\frac{\varphi^{}_T}{|\varphi^{}_T|} ~=~ e^{-i\alpha} \ .\label{eq:alpha}
\ee

Adopting this phase convention together with the alignments derived in
Section~\ref{sec:align}, we can deduce the mass matrices from the terms of the
lepton superpotential $W^{\mathrm{flavon}}_0$ of
Eq.~\eqref{eq:Wlepton1}. Mindful of the Clebsch-Gordan coefficients of $S_4$
in the $T$-diagonal basis, stated explicitly in Appendix~\ref{app:CGs}, we
obtain the Dirac neutrino mass matrix
\be
m^D ~=~ \frac{v_u}{\Lambda} 
\begin{pmatrix}
0 & \phantom{-}y^{}_{\mathrm{sol}}\,|\varphi^{}_{\mathrm{sol}}| \,\omega^{-k} \\
 -y^{}_{\mathrm{atm}}\varphi^{}_{\mathrm{atm}} &
-y^{}_{\mathrm{sol}}\,|\varphi^{}_{\mathrm{sol}}| \,\omega^{-k}   \\
\phantom{-}y^{}_{\mathrm{atm}}\varphi^{}_{\mathrm{atm}} &
\,3\,y^{}_{\mathrm{sol}}\,|\varphi^{}_{\mathrm{sol}}| \,\omega^{-k}  
\end{pmatrix},
\ee
where the only complex quantity is given explicitly by the factor
$\omega^{-k}$. Absorbing the minus signs into the second lepton doublet field,
which ultimately gets absorbed into the right-handed muon field when the
charged lepton masses are made real and positive, we obtain a Dirac mass
matrix with the sign conventions of Eq.~\eqref{mDn0} 
[also see Appendix~\ref{analytic}, Eq.~\eqref{mDn}]. 
This is our preferred convention which we will adopt in
the following. The $2\times 2$ right-handed Majorana mass matrix takes the
real and diagonal 
form $M_R=\mathrm{diag}(\xi^{}_{\mathrm{atm}},\xi^{}_{\mathrm{sol}})$,
continuing to suppress the dimensionless Yukawa couplings in the Majorana
sector for brevity. Applying the seesaw formula results in the effective light
neutrino mass matrix 
\be
	m^\nu = \frac{v_u^2}{\Lambda^2}\left[
\frac{ (y_{\mathrm{atm}} \varphi'_{\mathrm{atm}})^2}{\xi^{}_{\mathrm{atm}}}
	\begin{pmatrix}	0&0&0\\0&1&1\\0&1&1 \end{pmatrix} ~+~
	\frac{ (y_{\mathrm{sol}} |\varphi'_{\mathrm{sol}}|)^2}{\xi^{}_{\mathrm{sol}}}
\,\omega^{-2k} \,	\begin{pmatrix}	1&1&3\\1&1&3\\3&3&9  \end{pmatrix} \right] .
\ee
Choosing $k=2$, which is one of the three physically distinct possible choices $k=0,1,2$,
the neutrino mass matrix is of the form of Eq.~\eqref{eq:mnu2} but with 
fixed values of $n=3$ and $\eta = -2\pi/3$,
\begin{equation}
	m^\nu = m_a 
	\left(
\begin{array}{ccc}
	0&0&0\\0&1&1\\0&1&1 
	\end{array}
\right)
	+ m_b e^{-i 2\pi/3} 
	\left(
\begin{array}{ccc}
	1&1&3\\1&1&3\\3&3&9
	\end{array}
\right),
	\label{eq:mnu2p1}
\end{equation}
as in Eq.~\eqref{eq:mnu2p10} but with fixed phase $\eta = -2\pi/3$ 
leading to leptonic CP violation with Dirac phase $\delta \sim -\pi /2$, and good values 
of lepton mixing angles as discussed in Section~\ref{littlest}.
As the VEVs of all flavons which appear in the neutrino sector of
$W^{\mathrm{lepton}}_0$, see Eq.~\eqref{eq:Wlepton1}, respect the $SU$
symmetry, this neutrino matrix satisfies that symmetry as well and is therefore of the
trimaximal TM$_1$ form.

Considering the charged lepton sector, we find the diagonal mass matrix
\be
	m^\ell = \frac{v_d \, \varphi'_t}{\Lambda}
 \begin{pmatrix}	y'_e \frac{\varrho_t^2}{\Lambda^2}&0&0\\0&y'_\mu
   \frac{\varrho^{}_t}{\Lambda}  &0\\0&0&y'_\tau \end{pmatrix} 
   \equiv v_d
 \begin{pmatrix}	y_e &0&0\\0&y_\mu &0\\0&0&y_\tau \end{pmatrix}   ,
\ee
where the only complex quantity is given by the value of the doublet VEV
$\varrho^{}_t$. Its phase factor $e^{i\alpha}$ can, however, be absorbed into a
field redefinition of the right-handed electrons $E^c_2$ and $E^c_1$. As such,
the phase $\alpha$, as introduced in Eq.~\eqref{eq:alpha}, does not contribute
to the phase structure of the PMNS mixing matrix, and the effective Yukawa couplings
$y_e,y_\mu , y_\tau$ defined above may be taken to be real without loss of generality.
We emphasise that the hierarchy of physical effective Yukawa couplings
$y_e \ll y_\mu \ll  y_\tau  \ll 1$ has a natural explanation in this model,
arising from the smallness of flavon VEVs compared to the cut-off scale
$\Lambda$, assuming the primordial Yukawa couplings $y'_e,y'_\mu , y'_\tau \sim O(1)$.

As for the neutrinos, the structure of the charged lepton mass matrix $m^\ell$
is also related to symmetry, although in a slightly more intricate way. While the 
alignments of~$\phi'_t$ and~$\rho^{}_t$ do not change their direction under a
$T$ transformation, both pick up the phase factor
$\omega^2$.\footnote{For the triplet $\phi'_t$ this can be seen in
  Eq.~\eqref{eq:S4gens}. For the doublet $\rho^{}_t$ we note, that the
  corresponding $T$ generator is also diagonal with
  $T=\mathrm{diag}\,(\omega,\omega^2)$, see e.g.~\cite{Luhn:2013vna}.} A
subsequent $Z_3^{(3)}$ transformation $c^{(3)}$ can undo this change of the
phase so that the combined $c^{(3)}_{} T$ transformation can be identified as
the symmetry of the charged lepton sector which is responsible for guaranteeing
a diagonal mass matrix $m^\ell$. In this sense the model of leptons presented
here is a semi-direct model, since the residual symmetry of the lepton mass
matrices of both neutrino and charged lepton sectors may be identified as
different subgroups of $S_4$, namely $SU$ in the neutrino sector and $T$
(combined with a $Z_3^{(3)}$ transformation) in the charged lepton sector.

%%%%%%%%%%%%%%%%%%%%%%%%%%%%%%%%%%%%%%%%%%%%%
\section{Charged lepton flavour violation}
\label{CLFV}
Since the model is supersymmetric we can expect charged lepton flavour
violation in this model, due to one-loop diagrams involving sleptons,
neutralinos and charginos~\cite{Borzumati:1986qx,Hisano:1995cp,King:1998nv}. 
In the mass insertion approximation, the processes arise from having
off-diagonal slepton mass squared and trilinear matrices at low energies in
the super-CKM basis in which the charged lepton masses are diagonal. 
With flavour symmetry present, the high energy slepton mass squared and
trilinear matrices are controlled by the flavour symmetry and generally yield
only small off-diagonal entries. Unfortunately this is a rather delicate and
complex issue, with precise estimates depending on an expansion in flavon
fields, canonical normalisation and rotations to the super-CKM basis in which
the charged lepton masses are diagonal, followed by renormalisation group
running to low energies, along the lines of a recent analysis based on an
$SU(5)\times S_4 \times U(1)$ Grand Unified Theory (GUT) of
flavour~\cite{Dimou:2015yng}.

Ignoring the effects of the operator expansion, canonical normalisation and
super-CKM rotations (which are anyway highly suppressed in this model where
the charged lepton mass matrix is diagonal), the slepton mass squared and
trilinear matrices do not violate flavour at high energies, and the only
remaining effect arises from renormalisation group running. Then, using the
analytic results in~\cite{Blazek:2002wq}, we may make a simple 
estimate for the branching ratio of $\mu \rightarrow e \gamma$ as follows.
At leading order in a mass insertion
approximation~\cite{Borzumati:1986qx,Hisano:1995cp,King:1998nv} 
the branching fraction of $\mu \rightarrow e \gamma$ is given
by~\cite{Blazek:2002wq}: 
\beq
{\rm BR}(\mu \rightarrow e \gamma)\approx 
        \frac{\alpha^3}{G_F^2}
        f(M_2,\mu,m_{\tilde{\nu}}) 
        |m_{\tilde{L}_{21}}^2|^2  \tan ^2 \beta\ ,
    \label{eq:BR(li_to_lj)}
\eeq
where the off-diagonal slepton doublet mass squared is given 
in the leading log approximation (LLA) by 
\beq
m_{\tilde{L}_{21}}^{2(\mathrm{LLA})}
\approx -\frac{(3m_0^2+A_0^2)}{8\pi ^2}|b|^2\ln \frac{M_{\rm GUT}}{M_{\rm
    sol}} \ ,
\label{lla}
\eeq
and the remainder of the notation is fairly standard and given in \cite{Borzumati:1986qx,Hisano:1995cp,King:1998nv}.
In the present model leptogenesis fixes $M_{\rm sol}=4\times 10^{10}$ GeV
and the neutrino fit fixes $m_b=v_u^2|b|^2/M_{\rm sol}\sim 2.7$~meV, which
implies $|b|\sim 10^{-3}$. 
The smallness of the Yukawa coupling $b$ is due to its non-renormalisable origin
$b\sim \frac{\varphi'_{\rm sol}}{\Lambda}$. This contrasts with other semi-direct models 
such as those in \cite{Dimou:2015yng}
where the neutrino Yukawa couplings are $O(1)$, and implies that in this model,
charged lepton flavour violation such as $\mu \rightarrow e \gamma$ will be relatively highly suppressed,
at least according to our very simple estimate based on the assumptions above.

%%%%%%%%%%%%%%%%%%%%%%%%%%%%%%%%%%%%%%%%%%%%%
%%%%%%%%%%%%%%%%%%%%%%%%%%%%%%%%%%%%%%%%%%%%%

\section{Conclusions}
\label{conclusions}
In this paper, guided by the principles of minimality and symmetry, we have been led 
to a highly predictive theory of neutrino mass and lepton mixing
in which all CP phases are fixed and 
the neutrino masses and the entire lepton mixing matrix are determined by only
two real input mass parameters.
Starting from the
most elegant mechanism for the origin of neutrino mass, namely the seesaw
mechanism, we have focused on the most minimal version involving two
right-handed neutrinos.
Pursuing minimality, we were then led to
consider a two right-handed neutrino seesaw model with one texture zero and a
constrained form of Dirac mass matrix involving only two independent Dirac
masses with the structure of Eq.~\eqref{mDn0}, simply related to the CSD3
structure in Eq.~\eqref{mD} by $L_2\leftrightarrow L_3$. 
Our main achievement is 
to show that the new version of CSD3 can be obtained from symmetry arguments
based on $S_4$, working in the basis where the diagonal $T$ generator can enforce the
diagonality of the charged lepton mass matrix due to a residual $Z_3$
symmetry, while the preserved $S_4$ subgroup $SU$ in the neutrino sector with
a residual $Z_2$ symmetry is instrumental in enforcing TM$_1$ mixing.
The resulting scheme combines minimality with symmetry, leading to a high degree
of predictivity, where the predictions are protected from
higher order corrections by the full symmetry of the model.

We then proposed a realistic model of leptons, based on $S_4\times U(1)$
symmetry, with two right-handed neutrinos, where a straightforward $F$-term
vacuum alignment results in a neutrino mass matrix with the form of
Eq.~\eqref{eq:mnu2p10}. 
The relatively simple model corresponds to the left half of Table~\ref{tab:charges}
(to the left of the double vertical lines) in which the symmetry is only
$S_4\times U(1)$. However in order to achieve the phenomenologically desired 
phase of $\eta = -2\pi/3$ we were forced to extend the symmetries of the 
model (but not the particle content) to include a $(Z_3)^5$ symmetry
in the right half of Table~\ref{tab:charges}
(to the right of the double vertical lines). This enabled us to impose a CP symmetry,
then spontaneously break it in a controlled way, such that the phase is
constrained to be one of the cube roots of unity, however leaving no residual
CP symmetry in the charged lepton or neutrino sectors. With the phase 
chosen from the cube roots of unity to be $\eta=-2\pi/3$, all CP phases are
fixed and the baryon asymmetry of the universe then will determine the
lighter solar right-handed neutrino mass to be $M_{\rm sol}=4\times 10^{10}$ GeV.
The model predicts a normal neutrino mass hierarchy with $m_1=0$,
reactor angle $\theta_{13}=8.7^o$, solar angle $\theta_{12}=34^o$, atmospheric
angle $\theta_{23}=44^o$, and CP violating oscillation phase $\delta_{\rm
  CP}=-93^o$, depending on the fit of the model to the neutrino masses and
possible renormalisation group corrections. These predictions  will be tested soon.

%%%%%%%%%%%%%%%%%%%%%%%%%%%%%%%%%%%%%%%%%%%%%

%%%%%%%%%%%%%%%%%%%%%%%%%%%%%%%%%%%%%%%%%%%%%

%%%%%%%%%%%%%%%%%%%%%%%%%%%%%%%%%%%%%%%%%%%%%

\section*{Acknowledgements}

The authors are grateful to the Mainz Institute for Theoretical
Physics (MITP) for its hospitality and its partial support which enabled some of this work
to be performed.
SFK acknowledges support from the STFC Consolidated grant ST/L000296/1 and the
European Union Horizon 2020 research and innovation programme under the Marie 
Sklodowska-Curie grant agreements InvisiblesPlus RISE No. 690575 and 
Elusives ITN No. 674896.
The work of CL is supported by the Deutsche Forschungsgemeinschaft (DFG) within
the Research Unit FOR 1873 ``Quark Flavour Physics and Effective Field Theories''.

%%%%%%%%%%%%%%%%%%%%%%%%%%%%%%%%%%%%%%%%%%%%%

%%%%%%%%%%%%%%%%%%%%%%%%%%%%%%%%%%%%%%%%%%%%%

%%%%%%%%%%%%%%%%%%%%%%%%%%%%%%%%%%%%%%%%%%%%%

\appendix

\section*{Appendix}

%%%%%%%%%%%%%%%%%%%%%%%%%%%%%%%%%%%%%%%%%%%%%

\section{\label{app:CGs}$\boldsymbol{S_4}$ group theory }

Throughout this paper we work in the $T$ diagonal basis of $S_4$, as in \cite{King:2013eh}:
\be
S=\frac{1}{3}
\begin{pmatrix}
 -1 & 2 & 2 \\
 2 & -1 & 2 \\
2 & 2 & -1
\end{pmatrix},\ \ \ \ 
T=
\begin{pmatrix}
 1 & 0 & 0 \\
 0 & \omega^2 & 0 \\
0 & 0 & \omega
\end{pmatrix}
\ \ \ \ { \rm for} \ {\bf{3}} \   { \rm or} \  {\bf{3'}}  \ ,
\label{eq:S4gens}
\ee
and
\be
U=\mp
\begin{pmatrix}
 1 & 0 & 0 \\
 0 & 0 & 1 \\
0 & 1 & 0
\end{pmatrix},\ \ \ \ 
SU=US=\mp \frac{1}{3}
\begin{pmatrix}
 -1 & 2 & 2 \\
 2 &  2 & -1 \\
2  & -1 & 2
\end{pmatrix}, \ \ \ \ { \rm for } \ \ {\bf{3}}, {\bf{3'}} \ \  { \rm respectively. }
\ee

In this basis the symmetry preserving vacuum alignments are as follows:
$$
\phi_T \sim {\bf{3}} \sim  \begin{pmatrix}
 1 \\
 0\\
 0
\end{pmatrix},\   { \rm preserves  } \ T, \  { \rm breaks  } \ S,U, 
$$
$$
\phi_T' \sim {\bf{3'}} \sim  \begin{pmatrix}
 1 \\
 0\\
 0
\end{pmatrix},\  { \rm preserves  } \ T,U \  { \rm breaks  } \ S, 
$$
$$
\phi_S \sim {\bf{3}} \sim \begin{pmatrix}
 1 \\
 1\\
 1
\end{pmatrix},\  { \rm preserves  } \ S\  { \rm breaks  } \ T,U ,
$$
$$
\phi_S' \sim {\bf{3'}} \sim \begin{pmatrix}
 1 \\
 1\\
 1
\end{pmatrix},\  { \rm preserves  } \ S,U\  { \rm breaks  } \ T,
$$
$$
\phi_{SU} \sim {\bf{3}} \sim \begin{pmatrix}
 2 \\
 -1\\
 -1
\end{pmatrix},\  { \rm preserves  } \ SU\  { \rm breaks  } \ T,U ,
$$
and the two important $SU$ preserving alignments for ${\bf{3'}}$ flavons,
\be
 \phi'_{\rm atm} \sim {\bf{3'}} \sim \begin{pmatrix}
 0 \\
 1\\
 -1
\end{pmatrix},\  { \rm preserves  } \ SU\  { \rm breaks  } \ T,U ,
\ee
\be
 \phi'_{\rm sol} \sim {\bf{3'}} \sim \begin{pmatrix}
 1 \\
 n\\
 2-n
\end{pmatrix},\  { \rm preserves  } \ SU\  { \rm breaks  } \ T,U ,
\ee
where we fix $n=3$ such that
\be
 \phi'_{\rm sol} \sim {\bf{3'}} \sim \begin{pmatrix}
 1 \\
 3\\
 -1
\end{pmatrix},\  { \rm preserves  } \ SU\  { \rm breaks  } \ T,U .
\ee

In the following we summarise the Kronecker products and Clebsch-Gordan
coefficients. The non-trivial $S_4$ product rules are listed
below, where we use  the number of primes within the expression
\be
{\bs{\alpha}}^{(\prime)}  \otimes {\bs {\beta}}^{(\prime)} ~\rightarrow
~{\bs{\gamma}}^{(\prime)} \ , \label{CGnotation} 
\ee
to classify the results. We denote this number by $p$, e.g. in ${\bf 3}\otimes
{\bf 3}^\prime \rightarrow {\bf 3}^\prime$ we get $p=2$.
Then the Clebsch-Gordan coefficients are given as follows~\cite{King:2011zj}:
\\
$$
\begin{array}{lll}
{\bf 1}^{(\prime)} \otimes {\bf 1}^{(\prime)} ~\rightarrow ~{\bf
  1}^{(\prime)} ~~
\left\{ \begin{array}{c} 
~\\p=\mathrm{even}\\~
\end{array}\right.
&%
\left.
\begin{array}{c} 
{\bf 1}^{\phantom{\prime}} \otimes {\bf 1}^{\phantom{\prime}} ~\rightarrow ~{\bf 1}^{\phantom{\prime}}\\
{\bf 1}^{{\prime}} \otimes {\bf 1}^{{\prime}} ~\rightarrow ~{\bf 1}^{\phantom{\prime}}\\
{\bf 1}^{\phantom{\prime}} \otimes {\bf 1}^{{\prime}} ~\rightarrow ~{\bf 1}^{{\prime}}
\end{array}
\right\}
&
\alpha \beta \ ,
\\[10mm]
{\bf 1}^{(\prime)} \otimes \;{\bf 2} \;~\rightarrow \;~{\bf 2}^{\phantom{(\prime)}}~~ \left\{
\begin{array}{c}
p=\mathrm{even} \\
p=\mathrm{odd}
\end{array} \right.
&%
\left.
\begin{array}{c}
{\bf 1}^{\phantom{\prime}} \otimes {\bf 2} ~\rightarrow ~{\bf 2} \\
{\bf 1}^{\prime} \otimes {\bf 2} ~\rightarrow ~{\bf 2}\\
\end{array}\;~
\right\}
&
 \alpha \begin{pmatrix} \beta_1 \\ (-1)^p \beta_2\end{pmatrix}  ,
\\[7mm]
{\bf 1}^{(\prime)} \otimes {\bf 3}^{(\prime)} ~\rightarrow ~{\bf 3}^{(\prime)}
~~ \left\{ \begin{array}{c}
~\\[3mm]p=\mathrm{even} \\[3mm]~
\end{array}\right.
&%
\left. 
\begin{array}{c}
{\bf 1}^{\phantom{\prime}} \otimes {\bf 3}^{\phantom{\prime}} ~\rightarrow ~{\bf 3}^{\phantom{\prime}}
\\
{\bf 1}^{{\prime}} \otimes {\bf 3}^{{\prime}} ~\rightarrow ~{\bf 3}^{\phantom{\prime}}
\\
{\bf 1}^{\phantom{\prime}} \otimes {\bf 3}^{{\prime}} ~\rightarrow ~{\bf 3}^{{\prime}}
\\
{\bf 1}^{{\prime}} \otimes {\bf 3}^{\phantom{\prime}} ~\rightarrow ~{\bf 3}^{{\prime}}
\end{array}
\right\}
&
 \alpha   \begin{pmatrix} \beta_1 \\  \beta_2\\\beta_3 \end{pmatrix}  ,
\\[12.2mm]
{\bf 2} \;\; \otimes \;\;{\bf 2} \;~\rightarrow \;~{\bf 1}^{(\prime)} ~~ \left\{\begin{array}{c}
p=\mathrm{even}\\
p=\mathrm{odd}
\end{array}\right.
&%
\left.
\begin{array}{c}
{\bf 2} \otimes {\bf 2} ~\rightarrow ~{\bf 1}^{\phantom{\prime}} \\
{\bf 2} \otimes {\bf 2} ~\rightarrow ~{\bf 1}^{{\prime}} 
\end{array}~\;
\right\}
&
 \alpha_1 \beta_2 + (-1)^p \alpha_2 \beta_1 \ , 
\\[7mm]
{\bf 2} \;\;\otimes \;\; {\bf 2} ~\;\rightarrow \;~{\bf 2}^{\phantom{(\prime)}} ~~ \left\{ \begin{array}{c}
~\\[-3mm] p=\mathrm{even}\\[-3mm]~
\end{array}\right.
&%
\left.
\begin{array}{c}
~\\[-3mm]
{\bf 2} \otimes {\bf 2} ~\rightarrow ~{\bf 2} \\[-3mm]~
\end{array}~~\,
\right\}
&
   \begin{pmatrix} \alpha_2 \beta_2 \\  \alpha_1\beta_1 \end{pmatrix} , 
\\[7mm]
{\bf 2}\;\; \otimes \; {\bf 3}^{{(\prime)}} ~\rightarrow ~{\bf 3}^{{(\prime)}} ~~ \left\{\begin{array}{c}
~\\[-2mm] p=\mathrm{even}\\ \\[2mm]
p=\mathrm{odd}\\[-2mm]~
\end{array}\right.
&%
\left.
\begin{array}{c}
{\bf 2} \otimes {\bf 3}^{\phantom{\prime}} ~\rightarrow ~{\bf 3}^{\phantom{\prime}} \\
{\bf 2} \otimes {\bf 3}^{{\prime}} ~\rightarrow ~{\bf 3}^{{\prime}} \\[3mm]
{\bf 2} \otimes {\bf 3}^{\phantom{\prime}} ~\rightarrow ~{\bf 3}^{{\prime}} \\
{\bf 2} \otimes {\bf 3}^{{\prime}} ~\rightarrow ~{\bf 3}^{\phantom{\prime}} 
\end{array}\;
\right\}
&
 \alpha_1 \begin{pmatrix} \beta_2 \\  \beta_3\\\beta_1 \end{pmatrix} + (-1)^p
\alpha_2 \begin{pmatrix} \beta_3 \\  \beta_1\\\beta_2 \end{pmatrix}  ,
\\[13.5mm]
{\bf 3}^{(\prime)} \otimes {\bf 3}^{(\prime)} ~\rightarrow ~{\bf 1}^{(\prime)}
~~ \left\{ \begin{array}{c}
~\\p=\mathrm{even}\\~
\end{array}\right.
&%
\left.\begin{array}{c}
{\bf 3}^{\phantom{\prime}} \otimes {\bf 3}^{\phantom{\prime}} ~\rightarrow ~{\bf 1}^{\phantom{\prime}}
\\
{\bf 3}^{{\prime}} \otimes {\bf 3}^{{\prime}} ~\rightarrow ~{\bf 1}^{\phantom{\prime}}
\\
{\bf 3}^{\phantom{\prime}} \otimes {\bf 3}^{{\prime}} ~\rightarrow ~{\bf 1}^{{\prime}}
\end{array}\right\}
&
 \alpha_1 \beta_1 +\alpha_2\beta_3+\alpha_3\beta_2 \ ,
\\[9mm]
\end{array}
$$
$$
\begin{array}{lll}
{\bf 3}^{(\prime)} \otimes {\bf 3}^{(\prime)} ~\rightarrow ~{\bf 2}^{\phantom{(\prime)}} ~~
\left\{ \begin{array}{c}
~\\[-3mm]
p=\mathrm{even}\\ \\[1mm]
p=\mathrm{odd}\\[-4.5mm]~
\end{array}\right.
&%
\left.\begin{array}{c}
{\bf 3}^{\phantom{\prime}} \otimes {\bf 3}^{\phantom{\prime}} ~\rightarrow ~{\bf 2} \\
{\bf 3}^{{\prime}} \otimes {\bf 3}^{{\prime}} ~\rightarrow ~{\bf 2} \\[3mm]
{\bf 3}^{\phantom{\prime}} \otimes {\bf 3}^{{\prime}} ~\rightarrow ~{\bf 2} \\
\end{array}\;
\right\}
&
\begin{pmatrix} \alpha_2 \beta_2 +\alpha_3 \beta_1+\alpha_1\beta_3\\ 
(-1)^p(\alpha_3 \beta_3+\alpha_1\beta_2+\alpha_2\beta_1) \end{pmatrix} ,
\\[10.5mm]
{\bf 3}^{(\prime)} \otimes {\bf 3}^{(\prime)} ~\rightarrow ~{\bf 3}^{(\prime)}
~~ \left\{\begin{array}{c}
~\\p=\mathrm{odd}\\~
\end{array}\right.
&%
\left.\begin{array}{c}
{\bf 3}^{\phantom{\prime}} \otimes {\bf 3}^{\phantom{\prime}} ~\rightarrow ~{\bf 3}^{{\prime}}
\\
{\bf 3}^{\phantom{\prime}} \otimes {\bf 3}^{{\prime}} ~\rightarrow ~{\bf 3}^{\phantom{\prime}}
\\
{\bf 3}^{{\prime}} \otimes {\bf 3}^{{\prime}} ~\rightarrow ~{\bf 3}^{{\prime}}
\end{array}\right\}
& 
\begin{pmatrix} 
2 \alpha_1 \beta_1-\alpha_2\beta_3-\alpha_3\beta_2 \\  
2 \alpha_3 \beta_3-\alpha_1\beta_2-\alpha_2\beta_1 \\  
2 \alpha_2 \beta_2-\alpha_3\beta_1-\alpha_1\beta_3 
 \end{pmatrix} ,
\\[9mm]
{\bf 3}^{(\prime)} \otimes {\bf 3}^{(\prime)} ~\rightarrow ~{\bf 3}^{(\prime)}~~\left\{\begin{array}{c}
~\\p=\mathrm{even}\\~
\end{array}\right.
&%
\left.\begin{array}{c}
{\bf 3}^{\phantom{\prime}} \otimes {\bf 3}^{\phantom{\prime}} ~\rightarrow ~{\bf 3}^{\phantom{\prime}}
\\
{\bf 3}^{{\prime}} \otimes {\bf 3}^{{\prime}} ~\rightarrow ~{\bf 3}^{\phantom{\prime}}
\\
{\bf 3}^{\phantom{\prime}} \otimes {\bf 3}^{{\prime}} ~\rightarrow ~{\bf 3}^{{\prime}}
\end{array}\right\}
&
\begin{pmatrix} 
\alpha_2\beta_3-\alpha_3\beta_2 \\  
\alpha_1\beta_2-\alpha_2\beta_1 \\  
\alpha_3\beta_1-\alpha_1\beta_3 
 \end{pmatrix}  .
\end{array}\\[3mm]
$$

%%%%%%%%%%%%%%%%%%%%%%%%%%%%%%%%%%%%%

%%%%%%%%%%%%%%%%%%%%%%%%%%%%%%%%%%%%%

%%%%%%%%%%%%%%%%%%%%%%%%%%%%%%%%%%%%%

\section{A new type of CSD$\boldsymbol{n}$}
\label{analytic}

We may define a new general class of CSD$n$ models as follows.
In the diagonal charged lepton and two right-handed neutrino mass basis,
CSD$n$ is defined in this paper, up to phase choices, by the Dirac mass matrix
in LR convention:\footnote{Note that this version of CSD$n$ differs from that
considered in~\cite{King:2015dvf}, where the second column of the Dirac mass
matrix was $(b,nb,(n-2)b)^T$. For this reason we consider the TB mixing matrix
in a different convention. Compared to the analytic formulas
in~\cite{King:2015dvf}, the new version of CSD$n$ leads to a change in sign in
the parameters $y$ and hence $t$ and  $\epsilon^{\nu}$,
with $x,z,A,B$ unchanged, compared to the original version. 
This implies that the reactor and solar mixing angle formulas are unchanged,
but the atmospheric angle formula changes due to the sign change in
$\epsilon^{\nu}$, which has the effect of reversing the octant for the
atmospheric angle. The formula for $\sin \delta$ also involves a change in sign.}
\begin{equation}
	m^D = \pmatr{0 & b \\ a & (n-2)b \\ a & nb } .
	\label{mDn}
\end{equation}
The (diagonal) right-handed neutrino mass matrix $M_{R}$
with rows $(\overline{N^c}_{\rm atm}, \overline{N^c}_{\rm sol})^T$ and columns 
$(N_{\rm atm}, N_{\rm sol})$
is,
\begin{equation}
M_{R}=
\left( \begin{array}{cc}
M_{\rm atm} & 0 \\
0 & M_{\rm sol}
\end{array}
\right),\ \ \ \ 
M^{-1}_{R}=
\left( \begin{array}{cc}
M^{-1}_{\rm atm} & 0 \\
0 & M^{-1}_{\rm sol}
\end{array}
\right).
\label{mR}
\end{equation}
The low energy effective Majorana neutrino mass matrix is given by the seesaw formula
\beq
m^{\nu}=-m^DM_{R}^{-1}{m^D}^T,
\label{seesawp}
\eeq
which, after multiplying the matrices in Eqs.~(\ref{mDn},\ref{mR}), for 
a suitable choice of physically irrelevant overall phase, gives
\begin{equation}
	m^\nu = m_a 
	\left(
\begin{array}{ccc}
	0&0&0\\0&1&1\\0&1&1 
	\end{array}
\right)
	+ m_b e^{i\eta} 
	\left(
\begin{array}{ccc}
	1&n-2&n\\n-2&(n-2)^2&n(n-2)\\n&n(n-2)&n^2
	\end{array}
\right),
	\label{eq:mnu2}
\end{equation}
where $\eta$ is the only physically important phase, which depends on the
relative phase between the second and first column of the Dirac mass matrix,
$\arg (b/a)$, as well as $m_a=\frac{|a|^2}{M_{\rm atm}}$ and 
$m_b=\frac{|b|^2}{M_{\rm sol}}$. This can be thought of as the minimal (two
right-handed neutrino) predictive seesaw model since only four real parameters
$m_a, m_b, n, \eta$ describe the entire neutrino sector (three neutrino masses
as well as the PMNS matrix, in the diagonal charged lepton mass basis). $\eta$
is identified with the leptogenesis phase, while $m_b$ is identified with the
neutrinoless double beta decay parameter $m_{ee}$. 

Consider the tri-bimaximal TB mixing matrix~\cite{Harrison:2002er} in the
following sign convention: 
\begin{equation}\label{TB}
U_{\rm TB} =
\left(
\begin{array}{ccc}
\sqrt{\frac{2}{3}} &  \frac{1}{\sqrt{3}}
&  0 \\  \frac{1}{\sqrt{6}}  & -\frac{1}{\sqrt{3}} & - \frac{1}{\sqrt{2}} \\
-\frac{1}{\sqrt{6}} & \frac{1}{\sqrt{3}} &  -\frac{1}{\sqrt{2}}    
\end{array}
\right).
\end{equation}%
We then observe from Eq.~\eqref{eq:mnu2} that
\begin{equation}
m^\nu
\left(
\begin{array}{c}
2 \\
1\\
-1
\end{array}
\right)
=
\left(
\begin{array}{c}
0 \\
0\\
0
\end{array}
\right).
\label{CSD(n)a}
\end{equation}
In other words the column vector $(2,1,-1)^T$
is an eigenvector of $m^\nu $ with a zero eigenvalue, i.e. it is the first column of the TB mixing matrix,
corresponding to $m_1=0$. We conclude that the neutrino mass matrix 
leads to so-called TM$_1$ mixing~\cite{Xing:2006ms,Albright:2008rp},
in which the first column of the mixing matrix 
is fixed to be that of the TB mixing matrix, but the other two columns are not 
uniquely determined,
\begin{equation}\label{TM1}
U_{\rm TM1} =
\left(
\begin{array}{ccc}
\sqrt{\frac{2}{3}} & -
&  - \\ \frac{1}{\sqrt{6}}  & - & -  \\
-\frac{1}{\sqrt{6}} & - & -  
\end{array}
\right).
\end{equation}%

Since the neutrino mass matrix yields TM$_1$ mixing as discussed above, it can be block diagonalised by the TB mixing matrix,
\begin{equation}
	m^\nu_{\rm block} = U_{\rm TB}^T m^\nu U_{\rm TB} =
\left(
\begin{array}{ccc}
	0&0&0\\0&x & y\\0& y &
	z
	\end{array}
\right)	,
\label{eq:mnu3}
\end{equation}
where we find,
\beq
x= 3m_b e^{i\eta} , \ \ \ \ 
y=  -\sqrt{6}m_b e^{i\eta} (n-1),\ \ \ \ 
z= |z|e^{i\phi_z}= 2[ m_a+ m_b e^{i\eta} (n-1)^2 ]\ .
\label{xyz}
\eeq
It only remains to put $m^\nu_{\rm block} $ into diagonal form, with real positive masses,
which can be done exactly analytically of course,
since this is just effectively a two by two complex symmetric matrix
which may be diagonalised with a rotation angle $\theta_{23}^{\nu}$.
This procedure leads to the following exact analytic results for neutrino masses and lepton mixing parameters~\cite{King:2015dvf}.

Taking the Trace (T) and Determinant (D) of the non-trivial $2\times 2$
neutrino mass matrix times its Hermitian conjugate we find 
\bea
m_2^2+m_3^2 &=& T \equiv  |x|^2+ 2|y|^2 +|z|^2\ ,\\
m_2^2m_3^2 &=& D \equiv |x|^2|z|^2 +|y|^4   - 2 |x||y|^2|z|\cos A \ ,
\eea
from which we extract the exact results for the neutrino masses,
\bea
m_3^2 & = & \frac{1}{2}T +   \frac{1}{2} \sqrt{T^2-4D}\ ,\label{m3}\\
m_2^2 & =  & D / m_3^2\ , \label{m2}\\
m_1^2 & =  & 0 \ .\label{m1}
\eea

The exact expression for the reactor angle is given below,
\beq
\sin \theta_{13}  =  \frac{1}{\sqrt{6}}\left(1-\sqrt{\frac{1 }{1+t^2}}       \right)^{1/2},
\label{s13p}
\eeq
where 
\beq
t =   \frac{- 2 \sqrt{6}m_b  (n-1)}{2| m_a+ m_b e^{i\eta} (n-1)^2|\cos (A - B)
  -3m_b  \cos B}  \ ,
\label{tp}
\eeq
with
\beq
\tan B =   \frac{ 2| m_a+ m_b e^{i\eta} (n-1)^2|\sin A}{3m_b+2| m_a+ m_b
  e^{i\eta} (n-1)^2|\cos A}  \ ,
\label{Bp}
\eeq
and 
\beq
A= \arg [m_a+ m_b e^{i\eta} (n-1)^2 ] - \eta .\label{Ap}
\eeq

The solar angle is given in terms of the reactor angle by the TM$_1$ mixing sum rule in three equivalent exact forms,
\beq
\tan \theta_{12} = \frac{1}{\sqrt{2}}\sqrt{1-3s^2_{13}}\ \ \ \ {\rm or} \ \ \ \ 
\sin \theta_{12}= \frac{1}{\sqrt{3}}\frac{\sqrt{1-3s^2_{13}}}{c_{13}} \ \ \ \ {\rm or} \ \ \ \ 
\cos \theta_{12}= \sqrt{\frac{2}{3}}\frac{1}{c_{13}} \ ,
\label{t12p}
\eeq
where we have defined $s_{ij}=\sin\theta_{ij}$ and $c_{ij}=\cos\theta_{ij}$.
To first order in $s_{13}$, The solar angle $\tan \theta_{12}$ approximately
takes the TB value of $1/\sqrt{2}$.  

The exact expression for the atmospheric angle is given by
\beq
\tan \theta_{23} =\frac{|1+\epsilon^{\nu}_{23}|}{|1-\epsilon^{\nu}_{23}|}\ ,
\label{t23p}
\eeq
where 
\beq
\epsilon^{\nu}_{23} = \sqrt{\frac{2}{3}}t^{-1} 
\left[\sqrt{1+t^2 } -1 \right] e^{-iB} \ ,
\label{epsnup}
\eeq
and $t$ and $B$ are given in Eqs.~(\ref{tp},\ref{Bp},\ref{Ap}).
The atmospheric angle $\tan \theta_{23}$ is maximal when 
$B=\pm \pi/2$ since then $|1+\epsilon^{\nu}_{23}|$ is equal to $|1-\epsilon^{\nu}_{23}|$.

Mixing sum rules for TM$_1$ mixing can be expressed as an exact relation for
$\cos \delta$ in terms of the other lepton mixing angles~\cite{Albright:2008rp},
\beq
\cos \delta = - \frac{\cot
  2\theta_{23}(1-5s^2_{13})}{2\sqrt{2}s_{13}\sqrt{1-3s^2_{13}}}\ .
\label{TM1sum}
\eeq
Note that, for maximal atmospheric mixing, $\theta_{23}=\pi/4$, we see that 
$\cot 2\theta_{23}=0$ and therefore 
this sum rule predicts $\cos \delta =0$,
corresponding to maximal CP violation $\delta = \pm \pi/2$.
The prospects for testing the TM$_1$ atmospheric sum rules
Eqs.~(\ref{t12p},\ref{TM1sum}) in future neutrino facilities 
was discussed in~\cite{Ballett:2013wya}.

Using the Jarlskog invariant \cite{Jarlskog:1985ht} we find the exact relation~\cite{King:2015dvf}:
\beq
\sin \delta = \frac{ 24m_a^3m_b^3(n-1)\sin \eta }{m_3^2 m_2^2 \Delta m_{32}^2
  s_{12}c_{12}s_{13}c_{13}^2s_{23}c_{23}}\ .
\label{sdelta}
\eeq
Note the positive sign in Eq.~\eqref{sdelta}, which means that, for $n>1$, the sign of $\sin \delta $
takes the same value as the sign of $\sin \eta$, in the convention we use to 
write our neutrino mass matrix, namely  $- \frac{1}{2}\overline{\nu_L} m^{\nu} \nu_{L}^c$.
The above exact results for $\cos \delta$ and $\sin \delta$ completely fix the value of the Dirac oscillation phase 
$\delta$.

%%%%%%%%%%%%%%%%%%%%%%%%%%%%%%%%%%%%%%%%%%%%

%%%%%%%%%%%%%%%%%%%%%%%%%%%%%%%%%%%%%%%%%%%%

%%%%%%%%%%%%%%%%%%%%%%%%%%%%%%%%%%%%%%%%%%%%

\end{document}